\documentstyle[12pt]{article}
\global\arraycolsep=2pt 

\newcommand{\be}{\begin{eqnarray}}
\newcommand{\ee}{\end{eqnarray}}
\newcommand{\la}{\langle}
\newcommand{\ra}{\rangle}

\newcommand{\beq}{\begin{equation}}
\newcommand{\eeq}{\end{equation}}
\begin{document}
\begin{titlepage}
\begin{flushright}
 hep-ph/9601348\\
January 1996
\end{flushright}
\vspace{0.3cm}
\begin{center}
\Large\bf  
Effective Field Theories as Asymptotic Series: From QCD to Cosmology.
\end{center}

\vspace {0.3cm}

 \begin{center}    {\bf Ariel R. Zhitnitsky}\footnote{
  e-mail address:arz@physics.ubc.ca }
 \end{center}
\begin{center}
{\it Physics Department, University of British Columbia,
6224 Agricultural  Road, Vancouver, BC V6T 1Z1, Canada}
\end{center}
 \begin{abstract}
We present some generic arguments   demonstrating
that an effective Lagrangian $L_{eff}$
which,  by definition,  
contains operators $O^n$ of arbitrary dimensionality  
in general 
is not convergent, but rather  an asymptotic series. It means that
the behavior of the far  distant terms has a specific factorial dependence
$L_{eff}\sim \sum_n \frac{c_n O^n}{M^{n}},~c_n\sim n! ,~n\gg1$.
 
We discuss a few  apparently different problems, which however have
something in common--   the aforementioned  $n!-$ behavior:

1.Effective long -distance
theory describing the collective fields in QCD;

2.Effective Berry phase potential which is obtained
by integrating over the fast degrees of freedom. 
As is known, the Berry potential is associated
with induced local gauge symmetry and    might be relevant
for the compactification problem at the Planck scale.

3.Nonlocal Lagrangians introduced by Georgi\cite{Georgi}
for appropriate treatment of the effective field theories
without power expanding.

4.The so-called improved action in  
lattice field theory  where      
the new,  higher dimensional operators
have  been  introduced into the theory  in order   to reduce
the   lattice artifacts. 

5.Cosmological constant problem and vacuum expectation values
in gravity.

We discuss some applications 
 of this, seemingly pure academic phenomenon,
to various physical   problems with typical energies
from $1 GeV$ to the Plank scale.

\end{abstract}
\end{titlepage} 
\vskip 0.3cm 
\noindent  

\setcounter{page}{1}
\section{ Introduction}
Today it is widely believed that all of our present realistic field theories
are actually not fundamental, but effective theories. The standard model is presumably
what we get when we integrate out
modes of very high energy from some unknown theory,
and like any other effective field theory, its lagrangian
density contains terms of arbitrary dimensionality, though 
the terms in the Lagrangian density with dimensionality greater
than four are suppressed by 
negative powers of a very large mass  $M$. Even in QCD, for
the calculation of processes at a few $GeV$  we would use
an effective field theory with heavier quarks integrated out, and such an
effective theory necessarily involves terms in the Lagrangian
of unlimited dimensionality.

The basic idea behind effective field theories is that a 
physical process at energy $E \ll M $
can be described in terms of an expansion in $E/M$, see recent reviews
\cite{Georgi1},\cite{Manohar},\cite{Kaplan}.
 In this case
we can limit ourself by considering only a few first
leading terms and neglect  the rest. 
In this paper we discuss not  this standard formulation of the problems,
but rather, we are interested in 
the behavior of the coefficients of the very high dimensional operators
in the  expansion. We shall demonstrate that these coefficients
 $c_n$     grow  as  fast as a factorial $n!$ for sufficiently large
  $n$. 
Thus, the  series  under discussion   is not a convergent, but an asymptotic one.
Such a behavior rases   problems both of fundamental nature, concerning the
status of the expansion and of practical
importance, as to whether divergences can be associated with new
physical phenomena.
It means, first of all, that in order to make sense, 
such a theory should be defined by some
specific  prescription, for example, 
 by Borel transformation.

 Let us note, that our remark about the  factorial
dependence of the series for    large $n\gg1$
is an  absolutely irrelevant issue for the analysis of
   standard problems
when we are interested in the low energy limit only.
We have nothing
new to say about these issues. 

However, sometimes 
we need  to know the behavior of whole series   when
the distant terms in the  series might be important.
In this case the analysis of the large order terms in the expansion
has some physical meaning.

Such a situation may occur in a variety of different problems
as will be discussed in a more detail later in the text. Now 
let us mention 
that in general it occurs  when the energy scale $E$ is close
to $M$ or/and when two or more intermediate,
 not well separated scales,
come into the game\cite{Georgi}.

This letter is organized in the following way.
In the next section we argue, by analyzing a couple of 
examples, that the factorial behavior of the coefficients in front of  
  the high dimensional operators,   is a very general
property of   effective field theories
\footnote{The generality of this phenomenon can be compared
with the well known property of the  large order behavior
in perturbative series\cite{Large}.
As is known a variety of different 
field theories (gauge theories, in particular) exhibits
a factorial growth of the coefficients
in the perturbative expansion with respect to coupling constant.
This growth in  perturbative expansion  is very different from the
phenomenon we are discussing, where the factorial behavior
is related to high dimensional operators, and not to 
coupling constant expansion.
However,  in spite of the   
apparent  difference of these phenomena, actually 
they have some common
general origin. We shall discuss this connection later.}.
 
In the last section we discuss
some possible applications of the obtained results to
   different field theories  with very  different
scales (from QCD problems to the cosmological
constant problem). 
\section{Basic Examples.}
\subsection{Main Idea.}
We  begin our analysis with the following
 remark. An effective field theory can be considered as
an particular case of the more general idea of the Wilson operator
product expansion (OPE). 
It has been demonstrated recently \cite{Shifman}, that the
  OPE for some specific correlation functions
(heavy-light quark system $\bar{Q}q$) in QCD   
is an{\bf asymptotic}, and not a convergent series.
The general arguments of the paper\cite{Shifman} have been explicitly
tested in $QCD_2$ (where the 
 vacuum structure as well as the  spectrum
of the theory is known)
with the same conclusion concerning the asymptotic nature of OPE  
\cite{ARZ}.
In both cases the arguments were based on the dispersion relations and 
the general
properties of the  spectrum of the theory. However, the experience with
large order behavior in perturbative series\cite{Large} teaches us
that the factorial growth of the coefficients
 is of very general nature and it is not 
specific property of some Green functions.

Thus, we expect that the asymptotic nature of the OPE 
has a  much more general origin and it is not related to the
specific correlation functions, for which it was found
for the first time\cite{Shifman}.

To be more specific and in order to explain what is going on
with the effective  theory when we integrate out the heavy degrees of freedom,
let us consider QED with one heavy electron of mass $M$.
The effective  field theory  for photons can be obtained by integrating out 
the fermion degrees of freedom.
The most general solution of this problem is not known,
however in the case of a    specific (constant) external electric field $E$
the corresponding expression  for $L_{eff}$
is  known (see. e.g. the  textbook \cite{Zuber}). 
In order to find the OPE coefficients 
for the  high dimensional operators $E^n$ 
one can expand $L_{eff}$ in power of  $E$:
\be
\label{1}
L_{eff}=M^4\sum_nc_n(\frac{E}{M^2})^n.
\ee
 
Of course, the eq.(\ref{1}) is not the most general form, because
it does not contain all possible operators, in particular those
operators which would contain some terms with  derivatives
$\sim\partial_{\mu}E$.  .
Our goal now is to demonstrate that we  do have a factorial behavior 
already in this simple  case where we select only
some specific class of operators, namely those
$\sim E^n$. 

Our next    step is as   follows.
First of all we    shall find  an exact formula for the $n-$
dependence of  the coefficients $c_n$; secondly,
we give a 
qualitative explanation 
of  why 
  such a factorial  behavior takes place. Our argumentation will be so general in form
that  it will be
perfectly  clear that this phenomenon is very universal in nature.

The effective Lagrangian for the problem can be written in the following way
\cite{Zuber}:
\be
\label{2}
L_{eff}=\frac{1}{8\pi^2}\int^{\infty}_{0}
\frac{ds}{s^2}[Ecth(Es)-\frac{1}{s}]e^{-isM^2},
\ee
where we denote the external field $E$ together with its coupling constant $e$.
We expand this expression in $E$   using the  formula
\be
\label{3}
\frac{1}{e^x-1}=\sum_{k=0}^{\infty}B_k\frac{x^{k-1}}{k!}
\ee
where $B_k$ are Bernoulli numbers.
For large $k$ these numbers as is known exhibit     factorial growth:
\be
\label{4}
B_{2n}=2(-1)^{n+1}(2n)!\sum_{r=1}^{\infty}\frac{1}{(2\pi r)^{2n}}\sim
2(-1)^{n+1}(2n)!\frac{1}{(2\pi)^{2n}},~~n\gg1.
\ee
Thus, the coefficients $c_n$ in the OPE (\ref{1})
are factorially divergent for    large $n$:
\be
\label{5}
c_{2n}=\frac{1}{8\pi^2}2^{2n}B_{2n}\frac{(2n-3)!}{(2n)!}\sim (2n)!.
 \ee
In particular, for $n=2$ this formula reproduces
the well-known Euler-Heisenberg Effective lagrangian $L_{EH}$,
which is nothing but the  first nontrivial term in the series (\ref{1}): 
\be
\label{6}
  L_{EH}=\frac{2}{45M^4}(\frac{e^2}{4\pi})^2E^4,
\ee
We have    redefined the  coupling constant $e$ 
in this expression to present the formula in a standard way. 

Now, how one can  understand this factorial behavior (\ref{5})
in simple terms? We suggest the following   almost trivial
explanation which however is a very universal in nature.  

Let us   
look at the function $L_{eff}(z)$(\ref{1}) as an analytical function of the
complex variable
$z=E/M^2$ for which the standard dispersion relations hold.
 The factorial growth of the coefficients  in the real
part of   $L_{eff}(z)$ implies that the
corresponding imaginary part  has  a very  
 specific behavior $ImL_{eff}(z)\sim e^{-1/z}$
which follows from the dispersion relations: 
\be
\label{7}
f(z)\sim\sum_nf_nz^n~~
f_n\sim(a)^nn!\sim
\int\frac{dz'}{(z')^{n+2}}Imf(z')\longleftrightarrow
Imf(z')\sim e^{-\frac{a}{z'}}
 \ee
Here we have introduced an arbitrary analytical function $f(z)$ to be   
more general.

At the same time, an imaginary part of the amplitude, as
 is known,     is related to 
to a real physical process: the  pair- creation in the strong external field.
We have fairly good physical intuition of what kind of dependence on the field 
one could expect for such a  physical process. 
Namely, as we shall discuss later,
this process can be thought as a penetration through
a potential barrier in the quasi-classical approximation.
So, 
from a physical point of view we would expect that 
the $E-$ dependence should have the following form  $ImL_{eff}(E)\sim e^{-1/E}$.
As we shall see this is exactly the case
for our    QED example (\ref{1}) and
in a full agreement
with  what the dispersion relations (\ref{7}) tell us.

Now, we would like to present the  
   explicit   formula  for
the  probability of  pair creation  
in the constant electric field $E$. It is given by
(see e.g.\cite{Zuber}):
\be
\label{8}
w=-\frac{1}{4\pi^2}\int^{\infty}_{0}
\frac{ds}{s^2}[Ecth(Es)-\frac{1}{s}]Im(e^{-isM^2}).
\ee
The ``only" difference with the formula (\ref{2})
is the replacement $Re(e^{-isM^2})\Rightarrow Im(e^{-isM^2})$.
However, this replacement completely modifies the analytical structure. 
Indeed, the explicit calculation of the coefficients in
 the power expansion for imaginary part in the formula (\ref{8})   
leads to the following 
 integrals which are zero
$\int dz \sin(z)z^{2n-3}\sim\sin[(n-1)\pi]=0$.
Thus, the  imaginary part is not expandable at $E=0$ in agreement with
our  arguments about     a
 singular behavior at this point $\sim e^{-1/E}$.

Fortunately, a direct
 calculation\footnote{This integral can be reduced, in according to
Cauchy theorem, to the calculation of the contributions
from the poles of the  $cth z$ function.},
 without  using an  expansion in power of $E$ can  easily be
performed with the
following final result,
 explicitly demonstrating the $e^{-1/z}$ structure(see e.g.\cite{Zuber}):
\be
\label{9}
w=\frac{E^2}{4\pi^3}\sum_{n=1}^{n=\infty}\frac{1}{n^2}\exp(-\frac{nM^2\pi}{E})
\ee

A few comments are in order. First, the behavior $w(z)\sim e^{-1/z}$
 is exactly what we expected. It can be interpreted as  penetration through
a potential barrier in the quasi-classical approximation.
Indeed, the standard  formula for the ionization of a state
with bound energy $-V\sim2M$ 
and external field $E$ is proportional to
$$\sim\exp(-2\int dx\sqrt{2M(V-Ex)}\sim \exp(-\frac{const.M^2}{E})$$
which   qualitatively explains the exact result (\ref{9}).

We are not pretending here to have derived  new result in QED. All these classical
formulae have been  well known for a many years.
Rather, we wanted to explain, by analyzing  this
QED example, the main source of the $n!$ dependence in the Effective
lagrangian. The Effective lagrangian, by definition,
is a series of operators of arbitrary dimensions
constructed from the
light fields $E$.
 This is   presumably  
obtained from some underlying field theory by integrating out
the heavy fields of mass $M$. It is perfectly clear that the 
probability of the physical
creation of the  heavy  particles with mass   $M$ in external field $E$
is strongly suppressed $\sim \exp(-\frac{1}{E})$.
{\bf The dispersion relations thus  unambiguously imply that 
  the  coefficients in the real part
of the effective lagrangian are  factorialy large.}

We believe that this simple explanation is so universal in form
that it can be applied to almost   arbitrary nontrivial effective 
field theories leading to the same conclusion about factorial behavior.
We shall 
consider another explanation of the same phenomenon later in the text, but
now we would like to note that the relation
between imaginary and real parts of the amplitudes
of course is well known, and heavily used in particle physics.
In particular, the recent analysis of
the $n!$ behavior  
in  the perturbative $\alpha_s^n$ expansion 
  shows \cite{Zakharov},
that the physical multiparticle cross section ( the imaginary part)
is exponentially small. This   important result
is a simple consequence of   dispersion
relations similar to eq.(\ref{7}). 

We would like to come back to   formula (\ref{5}) 
  to explain this factorial behavior in the OPE one more time 
from an absolutely independent point of view.
Again, we use QED as  an example to demonstrate an idea,
 however, as we shall see,
the arguments which follow are much more general and  universal in nature.

As is known, almost all nontrivial field theories exhibit  
factorial growth of coefficients in the perturbative expansion
with respect to coupling constant
\cite{Large}\footnote{Do not confuse this perturbative expansion
with OPE and Effective lagrangian we are dealing with. These
series  are very different
in nature, but they both exhibit an factorial growth.}.
This factorial dependence can be understood as 
the rapid growth of the number of   Feynman graphs\footnote{ Here we
do not discuss  the so-called renormalons which give the same 
factorial dependence, but have a very different origin.}.

Now, how one can understand  the nature of the Wilson OPE
in terms of the Feynman graphs? As is known the computational recipe
of the coefficients in the OPE is simple: it is necessary to
separate large and small distance physics.
Large distance physics is presented by operators
of light fields; the small distance contribution is explicitly
calculated from the underlying field theory. Technically,  
in order to carry out this program, we cut the 
perturbative graphs in all
possible ways over the photon lines (in general case, a photon
field will be replaced by some light degrees of freedom).
These lines present the external light fields. They  are combined together
in the specific way to organize all possible operators. The coefficients
in front of these operators 
can be explicitly calculated and they are determined by  the 
small distance physics. 

From this technical explanation 
of the    calculation of the coefficients in the Wilson OPE it should be  clear,
that {\bf if the underlying theory possesses  factorial growth
in the perturbative expansion, the Effective lagrangian constructed
from this theory exhibits the same factorial behavior for the  high
dimensional operators.}
The moral of this argument is very simple:  the factorial growth
of the perturbative expansion in the underlying theory 
can not disappear without trace. 
  It will show up in the coefficients 
of the high dimensional operators in the Effective lagrangian
obtained from the underlying theory.

Having demonstrated the main result on factorial growth
of the coefficients (in an  Effective lagrangian ) as   universal phenomenon,
we would like to discuss a few more examples.
\subsection{Nonlocal Lagrangian.}

The main goal of this section is  the demonstration of the fact
that in general the so-called nonlocal lagrangians \cite{Kirzhnits},\cite{Georgi}
exhibit  the  same feature we have  been discussing in the previous section.
Namely, irrespective of  the ``smearing " prescription of the nonlocal
part of interaction, the corresponding Effective action, obtained in the
standard way, will exhibit the factorial growing coefficients for
the high dimensional operators. 

Before going into   details, let us recall a few general results
concerning   nonlocal lagrangians. First of all, 
we refer to the old  review paper \cite{Kirzhnits} on this subject
regarding  the motivations. The recent interest 
on this subject was renewed in ref.\cite{Georgi} where  
it was advocated that such a lagrangians is the useful tool
to deal with a  physical situation in which the scales are not 
well separated. Anyhow, our main interest at the moment is not 
a physical application, but rather, the demonstration of some
 universal property for such kind of system.
The next relevant remark concerning a  nonlocal lagrangian
is as follows: the nonlocal, lowest dimensional coupling 
constant (let us say, quartic)
in general case  can induce some changes in the coupling terms with larger number of 
fields (let say, six eight,..). Thus, we are forced to consider an
effective Lagrangian with operators with an arbitrary number of low energy fields.
To be more specific, we shall consider the  following effective lagrangian
for the scalar field $\phi$ discussed in ref.\cite{Georgi}:
\be
\label{10}
L_{int}=\sum_{r=1}^{\infty}G_{2r}\phi^{2r}
\ee
Here, $G_{2r}$ are some nonlocal functions which are analytical in the region
of definition, depend as a consequence of momentum conservation on $2r-1$
 linearly independent momenta, and may have dimensions proportional to 
some power of an 
implicit scale of nonlocality $\Lambda$. We assume in what follows 
that the nonlocal couplings  in the bare action (\ref{10}) are of order
of one ( we mean by this that there is no strong dependence 
on $r$, like $r!$ or so); we shall   demonstrate in this case that the
interactions will give  the factorial growing    coefficients 
for the  high  dimensional operators in the 
corresponding  Effective 
lagrangian  obtained from (\ref{10}).

We start our analysis from the well understood $\phi^4$ interaction.
The issue of whether the interaction is 
 local  or nonlocal   is not relevant
for the analysis of the large order behavior in Effective theory.
As we  have discussed  in the previous section, 
in order to calculate the coefficients for the  high dimensional operators,
we have to: a).calculate the number of graphs for the given order $n$,  
b). cut the internal lines to organize the operators of the maximal dimensions.
For $\phi^4$ theory, it is well known\cite{Lipatov},\cite{Large}, that 
the large order behavior of perturbative series is $n!$. When we cut lines
in order to produce the external operators, each cut gives two 
external $\phi^2$ fields. Thus, we get operator $\phi^{2n}$
in the corresponding Effective lagrangian with coefficient $n!$ in front of it
(or, what is the same, we expect the
following  behavior for the $n-$th term $L_{eff}^{(n)}$ in the
effective Lagrangian $L_{eff}^{(n)}\sim(\frac{n}{2})!\phi^n$).

 We would like to generalize  this result for the   bare
 action with arbitrary dimensions (\ref{10}).
In the course of these calculations we  shall reproduce the $(\frac{n}{2})!$
behavior mentioned above. We shall demonstrate also that the
essential result    will not be changed with the increasing of  dimensions
 of the   vertices $r$
(\ref{10}) provided  that $n\gg r$. The last condition is required for the  method to be
applicable.

Let us remind that the Lipatov's idea \cite{Lipatov},\cite{Large}
of the calculation of large order behavior in a field theory
   is to present the coefficients
$Z_k $
in the perturbative expansion $Z(g)=\sum Z_kg^k$
 through a contour integral in the complex $g -$ plane:
\be
\label{11}
 Z_k\sim \int D\phi\oint\frac{dg}{g^{k+1}}e^{-S(\phi)},
\ee
where $S(\phi)$ is the action of the scalar field theory $\frac{g}{4}\phi^4$
and $D\phi$ is the standard measure for the functional integral which 
defines the theory (We discuss 
 here the perturbative expansion for the  Grand Partition Function
$Z(g)$. An arbitrary correlation function can be considered in  an 
analogous way. ). If the theory possesses the classical
instanton solution, then the calculation of the integral over $g$
can be done through steepest descent method.
 This method is
justified only for small $g$. But for  the large order $k$, the 
integral over $g$ is dominated by the small $g$ contribution.
Indeed,  in our specific case of  $\phi^4$ field theory
the classical instanton solution has the  following 
property $\phi_{cl}\sim 1/\sqrt{g}$,
\cite{Lipatov}. This can be seen from the saddle point equations
for $g_0(k)$ and $\phi_{cl}(k)$
(the actual equations are differential equations, of course, but 
we  are keeping the track only on external parameter $k$ ,
disregarding all complications  related to the coordinate $x$ dependence) :
\be
\label{12}
\frac{k}{g_0}+\frac{\phi_{cl}^4}{4}=0,~~
-\Delta\phi_{cl}+g_0\phi_{cl}^3=0,~~
\Longrightarrow\\ \nonumber
 \phi_{cl}\sim\frac{1}{\sqrt{g_0}}\sim\sqrt{k},~~
g_0\sim\frac{-1}{k},~~S_{cl}\sim\phi^2_{cl}\sim\frac{1}{g_0}\sim k.
\ee

From these equations it is clearly seen that
 the classical action $S_{cl}\sim 1/g_0\sim k$ is parametrically
large for  the large external
parameter $k$. Thus, the semiclassical approximation is completely
justified.

The generalization of these formulae for the more complicated
action $\sim \phi^{2r}$ is straightforward: Instead of 
(\ref{12}) we have the following behavior:
\be
\label{13}
 \phi_{cl} \sim\sqrt{kr},~~
g_0\sim\frac{-1}{k^{r-1}},~~S_{cl}\sim\phi^2_{cl} \sim k.
\ee
Thus, the method is applicable for the large $k$ and for any finite number $r$,
where the classical action is large and the coupling constant is small.
 From these formulae
one can calculate the   the large order behavior in perturbative 
series with bare action $\sim g\phi^{2r}$. The result is
$g^{k}(rk-k)!$. This growth is much faster than we found previously
for $\phi^4$ theory with $r=2$. However, 
the coefficients in the Effective lagrangian for the operator
$\phi^n$ grow in the same way as before $\sim (\frac{n}{2})!$.
The technical explanation for that is simple: when we cut 
the lines    in order to produce an external operator, the dimension
of the obtained operator $\sim\phi^{2k(r-1)}$ would be higher  than
for $\phi^4$ theory.
Thus, for the operator $\phi^n$ the coefficients in Effective lagrangian
have the same growth $\sim (\frac{n}{2})!$ as we already mentioned.

It would be interesting to understand this result in somewhat different way.
Essentially, what we need to calculate is the number of graphs which contribute
to the  $n$ point correlation function 
$Z^{(n)}\sim\la\phi(x_1)\phi(x_2)...\phi(x_n)\ra$.
Such a calculation can be done within the same Lipatov's 
technique. The only technical difference in comparison with the calculation
of the large order behavior for  the Grand Partition Function
$Z(g)$ itself is following:  We have to substitute   in the first approximation
the classical solution $\phi_{cl}$ in place of the 
external $\phi$ fields.
More precisely,
\be
\label{14}
 Z^{(n)}\sim \int D\phi\oint\frac{dg}{g}
  \phi(x_1)\phi(x_2)...\phi(x_n)  e^{-S(\phi)}\sim 
\ee
\be
\int D\phi\oint\frac{dg}{g}e^{-S(\phi_{cl})}
\phi_{cl}(x_1)\phi_{cl}(x_2)...\phi_{cl}(x_n) 
 \sim (\sqrt{n})^n\sim( \frac{n}{2})!
\nonumber
\ee
In this formula
we took into account that the classical field depends on $n$ as 
$\phi_{cl}\sim  \sqrt{n}$, (\ref{13}) and the total number of external fields
in the correlation function
is equal $n$.  The semiclassical approximation we have used in the 
derivation  (\ref{14})
is justified as far as number $(\frac{n}{2}) \gg 1$.
Only in this case the integral over $g$ is dominated by the small
$g$ contribution and instanton calculus can be applied.

 The  factorial dependence (\ref{14})
can be interpreted  as the rapid growth of the number
of Feynman graphs. As we see the dependence on
$n$ remains the same irrespectively to the form 
of the bare vortices provided that $n\gg r$.
This is in agreement with what we discussed before 
and related to the fact that the  essential part of classical solution
$\phi_{cl}\sim \sqrt{n}$ remains  the same for arbitrary $r$.
Such a behavior suggests that all terms
from the bare action give more or less the same contributions
 to the coefficients in the 
Effective theory.
 To obtain the total number of graphs wich contribute to the operator
$\phi^n$ we should sum up all terms coming from all possible  vertices $r\ll n$.
It gives essentially the same $(n/2)!$ behavior because
$\sum_{r=2}^{r\ll n}c_r r^{n/2}(n/2)!\sim(n/2)! ~~n\gg1$.
We do not expect any special cancellations between different terms
which may kill this growth.
The contribution from the higher order operators $r\simeq n$ can not be estimated
in the same way, but one could expect that the growth of
the coefficients could be even more severe in this case.

The moral is: We certainly have a divergent series for Effective lagrangian
induced by some unknown full theory no matter what the starting point is.
We shall discuss some applications of this result in the conclusion.
\subsection{ A few more examples}
In this subsection
we are going to discuss a few more examples
from   very different fields of physics:\\
a).
Collective fields in QCD;\\
 b). Berry phase as a dynamical
field in compactification problem;\\ c). Lattice field theories.\\
d)Gravity at Plank scale.

 We shall demonstrate that the phenomenon of the asymptotic nature
of an Effective lagrangian is a  very universal one. This universality is
the common feature which characterizes these so different fields of physics
we  mentioned above.
{\bf a).}
We start from the QCD,   as underlying theory.
The problem in this case can be formulated in the following way
(see recent paper \cite{Nielsen} on this subject and refences therein).
How one can integrate over small distance physics in order to
extract the long-distance dynamics? An appropriate way 
to implement this programm is: a). 
introduce  the collective degrees of freedom, colorless mesons,
 as the external sources
into the underlying lagrangian; b)integrate over the
quarks and gluons with high frequencies by introducing
the normalization point      $\mu$.
The obtained  Effective lagrangian is the $1/\mu$ expansion
where operators are expressed in terms of
  the external fields as well as 
low-energetic  quarks and gluons. Our remark is:  the coefficients
in this expansion grow  factorially   with the dimension of the operators.
We postpone the discussion of the  physical meaning of this result 
to the Conclusion. Let us note, that the procedure
of   obtaining   the Effective lagrangian in this case is
not much different from the case we discussed previously.
The only new element  is the introduction of the collective  fields
which were not present in our original lagrangian. However,
this does not effect the general arguments on the $n!$ behavior.

Indeed,  one can consider the
quark-antiquark external lines 
(instead of the  collective meson fields)
for the calculation
of the coefficients in the OPE, as discussed in the previous section.
In this case, all arguments on $n!$ behavior can be applied
in a straightforward way. Thus, {\bf we expect a factorial behavior
of the coefficients for the Effective QCD lagrangian,
as well as for   the chiral lagrangian, as its particular case.} 
An exact formula for the coefficients
depends on the operator under consideration.
This is because the  different fields (gluons, quarks, mesons), 
which are constituents of  the operator are 
not equally weighted. However the precise expression for the
coefficients in terms of constituents of 
these operators is not a relevant issue  at the moment.
We shall discuss consequences of this result in the Conclusion.

{\bf b).}
We continue our short review of different models
by analyzing the so-called 
Berry phase as a dynamical gauge field\cite {Wilczek}. There are a few applications
of this idea. We consider only one of them.
As is known, the standard philosophy of    compactification
at the Plank scale is the assumption of a very high gauge invariance
at this scale 
which will be broken at   lower scales. It is quite possible
that some of gauge symmetries are dynamically induced rather than   a 
required principle. 
We refer to the recent paper \cite{Kikkawa}
on this subject for details and references. Here we would like
to demonstrate that the Effective lagrangian for the induced
dynamical Berry field is not a convergent, but an asymptotic series.
As usual, the Effective lagrangian is obtained by integrating
over the fast degrees of freedom; the Berry field itself
is considered as a slow variable. The Effective lagrangian 
is understood as a theory describing the dynamics of
these slow fields. 

To be more specific,  if one    integrates over the compactified space
coordinates,  than one obtains an
Effective lagrangian which depends on Berry's potential
 $A_{\mu}=-iu^{\dagger}\partial_{\mu}u$. Here $u$ is an original fermion field
considered as a fast variable. Now the situation   clearly resembles
  QCD where  the underlying lagrangian does not contain   meson fields.
They will appear and
become dynamical variables after   integrating over the  fast quark fields.
The same situation takes place 
in the case under consideration   
 where the Berry potential can be thought of as a composite
of $u$ and $u^{\dagger}$ original fields. 

Now all previous QCD- arguments regarding the 
$n!$ growth of coefficients in the Effective lagrangian can be applied
to the present case. We end up with the same {\bf conclusion 
that the  Effective lagrangian $L_{eff}(A_{\mu})$ as a function of the Berry potential
is an asymptotic series}\footnote{The very different approach\cite{Berry}
 leads assentially to the similar conclusion
about the asymptotic nature of the adiabatic expansion}.
 Of course, there is a huge difference
in scales between QCD and the theory under consideration:
in former case the parameter of expansion is $1/\mu$ with
$\mu\simeq 1GeV$; in later one the scale is the Plank scale
$M_P\simeq 10^{19}GeV$. However, there is no fundamental difference
between these two models in the way of   obtaining
   the corresponding  Effective lagrangians: in both cases the slow fields
can be considered as the composite of the original fields.
A symmetry prevents them from getting  a mass:
for the $\pi$ meson it is the chiral symmetry; for the
Berry field $A_{\mu}$  it is a gauge symmetry.
Thus, both fields can be considered as   soft variables
and the philosophy of Effective lagrangian can be applied.
The integration over the fast variables, as we argued earlier,
leads to the $n!$ growth of the  coefficients in the Effective lagrangians
in both cases.

{\bf c).}
Our next example is the lattice QCD.
As is known, the main idea in lattice QCD is to replace continuous spacetime
variables by a discrete lattice. Then the path integral defining the QCD can be
 evaluated numerically. If we denote $a$ as the lattice spacing, then the
standard discretization of the QCD action has errors of $O(a^2)$
that are large when the lattice spacing
is not small enough. This was the reason to suggest the so-called
improved action for lattice QCD\cite{Symanzik}(for   recent development
see \cite{Lepage}).
 The improved discretization 
has been designed in such a way that finite $a$-errors are 
systematically removed by introducing new (nonrenormalizable) 
interactions into the lattice action. All coefficients of the new 
interactions are determined by demanding that the discretized
action reproduces continuum physics to a given accuracy.
In particular, the Wilson action contains all terms proportional
to $a^2, a^4,...$ beyond the desired gluon kinetic
term\cite{Symanzik}:
\be
\label{15}
1-\frac{1}{3}Re~TrU_{pl}=r_{0}\sum_{\mu\nu}Tr(F_{\mu\nu}F_{\mu\nu})
+a^2\sum_ir_{i}R^i+
\ee
\be
0(a^4)+...  
+\sum_{n,i} a^{2n}r_{ i, 2n}Q^{i,2n},\nonumber
\ee
where  $U_{pl}$ is the product of link matrices on a  plaquette $P$;
$R^i$ is the set of operators of dimension six; the $r_{i}$ are coefficients in the 
OPE of the plaquette. For   higher dimensional operators we introduced the
corresponding notations $ Q^{i,2n}$ and $r_{i, 2n}$   with the index $n$
labeling the  dimension of the operator, and the index $i$ classifying different
operators with given dimension.

Our remark is: {\bf The coefficients in the 
expansion (\ref{15}) are factorially  growing   with  the dimension
of the operators.} We shall discuss the  physical consequences
 of this statement in the Conclusion. Now, we would like to explain
this $n!$ growth in the following way:
The lattice action is defined in terms of the link operator
\be
\label{16}
U_{x,\mu}=P\exp[-ig\int_x^{x+a\mu} A\cdot dy]
\ee
with the simplest choice of path for the integral as a straight line joining
$x$ and $x+a\mu$.
 A single plaquette contribution can be thought of as a Wilson loop
surrounding this point $x$ with radius $a$. As is known, the Wilson loop
can be interpreted as the creation of a heavy quark-antiquark  pair
which
propagates for a time $a$ and finally   annihilates. It can be interpreted as
a forward and backward propagating of   one heavy quark as well.
Anyhow, one can interpret the action (\ref{15}) as the effective action
which is obtained after
integrating  out the heavy quarks with mass $a^{-1}$. As usual,  
to give some sense to the Effective lagrangian which   presumably
describes the dynamics of light degrees of freedom, the mass of the
auxiliary heavy quark should be much larger than the characteristic scale in QCD:
$a^{-1}\gg 1GeV$.
Once this interpretation in terms of the heavy quark has been made,
we have  reduced our problem to the previously discussed case (\ref{1}).

{\bf d).}
Our last, but not least example is the effective field theory of gravity.
We refer to the recent review \cite{Donoghue}
on this subject for a general introduction and references.
The only remark we would like to make here is the following.
Nowdays it is generally accepted
that the Einstein lagrangian
\be
\label{17}
S_{grav}=\int d^4x\sqrt{g}\frac{2}{\kappa^2}R
\ee
is  only the   first local  term of the expansion of a more complicated
theory (string?).
Thus,   general relativity should be considered as 
an effective  field theory 
with infinitely many terms allowed by general coordinate invariance.
As usual, in the effective theory description,  
 only the first term in the expansion plays a role at low energy
$E\ll M_{Plank}$. If we were not  interested in   
quantum effects at the Plank scale
with $E\simeq M_{Plank}$,   eq. (\ref{17}) would be the end of the
story.
However we intend to discuss   physics   at the Plank scale, 
thus we would like to write down the
Effective lagrangian in the most general form:
\be
\label{18}
S_{eff}=\int d^4x\sqrt{g}[\Lambda+\frac{2}{\kappa^2}R
+c_1R^2+c_2R_{\mu\nu}R^{\mu\nu}+ \\     \nonumber
\sum_{n}c_nQ^n...+L_{matter}+L_{dilaton}+L_{inflaton}...+],
\ee
where the operators $Q^n$ are high dimensional operators
constructed from the relevant fields ($R_{\mu\nu}$, dilaton, inflaton $\phi$,
gauge fields $F_{\mu\nu}$, etc).
Our remark here is  that {\bf  
the coefficients in the Effective lagrangian describing
 even the pure gravity theory,
exhibit  factorial growth.}\footnote{Any extra fields
may only  increase this growth.}
The arguments which support this statement are the same as before:
 if the underlying theory
(in our case it is given by
lagrangian (\ref{17}) possesses   factorial growth
in the perturbative expansion, the Effective lagrangian constructed
from this theory exhibits the same factorial behavior for the  high
dimensional operators.

 As we already mentioned, the factorial behavior of coefficients
in the perturbative expansion can be understood as the fast increase
in the number of Feynman diagrams. In pure Yang Mills
theory we know well that such a growth does take place\cite{Large}.
We can interpret this growth as  a manifestation of the 
three- and four- gluon vertices which lead to the factorially
divergent number of the diagrams. In the case
of gravity (\ref{17})   we expect the same factorial
behavior because of the nonlinear nature of the interaction
(\ref{17}) similar to a gauge theory.
Of course, there is a 
big difference between those two, related to the fact 
that   gravity is not a renormalizable field theory.
 However the only 
relevant point for our purposes    is that the coefficients are
factorially growing  the  dimension of the operator increases.
  The possible physical consequences of this phenomenon
will be discussed in the last section.

\section{Instead of conclusion}
\subsection{General summary}
In this letter we have presented two independent sets of arguments
which support the idea that almost any nontrivial Effective
lagrangian obtained by integrating out some
heavy fields and/or fast degrees of freedom,
is non-convergent, but an asymptotic series.

The first set of arguments is
based on the idea that the imaginary part of the amplitude   related to
  the probability of the physical creation of a heavy particle,
is exponentially small $\sim\exp(-\frac{1}{E})$. The dispersion relations
in this case unambiguously imply that the coefficients of the
expansion in the  real
part of the corresponding
amplitude exhibit an factorial dependence.
Once these coefficients are found to be factorially large,
we can forget about the way the result  was derived, we can forget 
about the external auxiliary field $E$ which we heavily used in our arguments.
Coefficients in the OPE do not depend on the applied field $E$,
no matter how small it is.

The second line of reasoning is  based on the analysis 
of the large order behavior of the  perturbative series.
As we have argued, if the underlying theory
possesses   factorial growth of the coefficients of the  perturbative series,
than the corresponding Effective lagrangian constructed from
this theory will exhibit the same factorial behavior
for the high dimensional operators. 

We believe that both of these lines of arguments are so general 
in form that almost all nontrivial Effective lagrangian
will demonstrate $n!$ behavior. We believe that this phenomenon is
   universal in nature.

Now we would like to discuss some physical consequences
which might result from this  phenomenon.
As we mentioned in  Introduction,
we have nothing new to say in the case of analysis of    
    low energy phenomena   for which 
 the small  expansion  parameter  is $\lambda\equiv E/M\ll1$.
In such a  case, the exact formula is approximated 
perfectly well by the
first term of the asymptotic expansion and we can safely forget
about all the rest.
However, very often  
the situation is not so 
fortunate and the expansion parameter      $\lambda \sim  1$,
(let say $1/3 $ or $1/2$).
In this event people try to improve the situation
by considering the next to -leading terms or even next to next 
to -leading order.
 If the series were convergent,
 these efforts would be worthwhile.
However, as we argued in this letter, an Effective lagrangian , in general,
is represented by an asymptotic, not a  convergent series.
Thus, one may ask the following general questions:\\
a).How many terms one should keep in the Effective lagrangian
for the best approximation of  an exact formula for the given 
parameter $\lambda$?\\
b).What is the {\bf fundamental } uncertainty (related to our lack 
of knowledge of the higher dimensional operators)
 one should expect for an Effective lagrangian
represented as an asymptotic series?\\

Let us recall that the standard perturbative expansion in QCD is also
asymptotic series. For this case the answers on the questions
a). and b). are well known\cite{Zakharov}.
In particular, as is known, the pole mass of a heavy quark
suffers from an intrinsic uncertainty of order $\Lambda_{QCD}$.
\cite{Shifman}
Another example is the fundamental uncertainty 
of perturbative calculations of the correlation
function for the light quarks\cite{Zakharov}.

We believe that the asymptotic nature of the OPE and   Effective lagrangian,
in particular,  will lead to a similar fundamental uncertainty
for some physically interesting characteristics.
In particular, as we argued in \cite{ARZ},
   any hopes to improve the standard QCD sum rules
(like the idea advocated in\cite{Raduyshkin}) by summing up
a certain subset of the power corrections  and ignoring all the rest,
is fundamentally an erroneous idea because of the asymptotic nature of the OPE.
A  similar example which has been discussed 
recently is the OPE for $\tau$ decay\cite{Shifman}.
It was argued that the tail
of the condensate series may be quite noticeable in the 
nonperturbative analysis of
the hadronic $\tau$ decay.

Therefore, the moral is: if the  parameter of the asymptotic expansion
is not small enough, the two questions formulated above
might  have some phenomenological relevance. The effective description of QCD  
which has been discussed in the previous section is
  one   example. 
We believe that the lattice calculations 
(also discussed in the previous section ) is another example 
of the same kind. Indeed, as we argued in the previous section
the expansion (\ref{15}) is an asymptotic series. Thus, we can formulate
 the following question: {\bf How many terms
in the asymptotic expansion should be kept 
for the  given lattice size $a$ in order to get the best
possible accuracy? } The same question can be reformulated in 
somewhat different way: What is the fundamental uncertainty
of the lattice calculations which are
associated with the tail of the high dimensional operators
in the Effective lagrangian (\ref{15})?
\subsection{Cosmological constant problem}
We wish to discuss some consequences of the factorial behavior 
in the    Effective lagrangian (\ref{18}) for gravity  separately.
Let us recall that the natural scale of the cosmological term
$\Lambda$ is the Plank scale. Indeed, the most popular cosmology today,
the inflationary scenario (for a review see
 \cite{Linde} and \cite{Robert})
 assumes that our universe passed through an era in which
the cosmological term dominated, and it is a total mystery why we
should be left in a universe with an almost vanishing vacuum energy.
Of course we do not know the answer to this question, but we 
would like to suggest the following scenario which is based
on the asymptotic nature of the effective lagrangian (\ref{18}).

Let us assume that at the very early epoch the gravity field as well as
other relevant  for inflation fields (scalar,..) 
  exhibit some nonzero vacuum expectation values (VEV), 
which we shall call the condensates.
We believe that this is very likely to happen in gravity at the Plank scale
in analogy with the  phenomenon of   
 gluon condensation in QCD at $1 GeV$ scale.
We introduce the  notation $\la  \Phi \ra$ for 
the condensate of any relevant  field: a  scalar field which
people usually introduce to describe inflation (inflaton), dilaton or a 
gravity field itself.
 The natural scale for such a condensate
is, of course, the  Plank scale. For the higher dimensional
operators $Q^n$ from   eq.(\ref{18}) we assume 
that there is a   factorization rule which 
allows us to estimate the higher order condensates in the following way
$\la Q^n \ra \sim \la \Phi^n \ra \sim \la \Phi \ra^n$.
We note that this assumption is not crucial for our purposes,
but, rather, is    a simplification which
allows us to demonstrate the main idea in a very simple way. 
The similar assumption in QCD is justified in the limit in which
the   number of colors
$N_c\rightarrow\infty$.Given that these assumptions have been made,
we can use the Borel representation formula for the
    asymptotic series (\ref{18})\footnote{
We assumed in this formula that the series is Borel summable.
This is may or  may not be the case; however we believe that the
Borel non-summability of an expansion does not signal
an inconsistency or ambigiuty of the theory.
The Borel prescription is just one of many summation methods
and need not be applicable everywhere. For Borel-non-sumable
cases, one could expect the sign $(-)$ in the denominator of eq.(\ref{20}).
Thus, some prescription, based on the physics consideration, should be given
in order to evaluate an integral like that. Some new physics
usualy accompanies such a phenomenon, but we do not go into details here.
Rather, we would like to
  mention the non-Borel sumable  example of the principal
chiral field theory at large N\cite{Fateev}. In this case,
the explicit solution is known. The coefficients
grow factorially with the order and the series is non-Borel sumable.
Nevertheless, the physical observables are perfectly exist, 
the exact result can be recovered by special prescription which
uses a non-trivial procedure of analytical continuation.}:
\be
\label{20}
\la L_{eff}\ra \sim\sum_{n=0}^{n=\infty}n!(-1)^n \la\Phi\ra^n=
\int_{0}^{\infty}\frac{dt}{t(t+\la\Phi\ra)}
\exp(-\frac{1}{t}) 
\ee

Now we would like to briefly discuss  the vacuum structure of
de Sitter Space. In different words, we would like to
discuss the parameter $\la\Phi\ra$ from   eq. (\ref{20}).
We refer to the recent papers \cite{Dolgov}-\cite{Woodard}
on this subject (see references to previous papers therein).
The main result of these investigations is the observation
that the higher order quantum gravity corrections to the different
physical values in general are infrared divergent.
In particular,  
the divergence is observed in the
 vacuum correlator $ \la\phi^2\ra $.  
 Probably, this divergence 
has power-like behavior in time rather than exponential one, as previously thought.
It may force us to take some nonperturbative dynamics into account,
which we  do not know.
Instead we introduce   some small phenomenological parameter $\epsilon(\tau)$  
into the 
VEV  $$ \la\Phi\ra \rightarrow\frac{\la\Phi\ra}{\epsilon(\tau)
},~~\epsilon\rightarrow 0$$
in order to account for this new physics
responsible for the infrared divergences mentioned above.

One can see in this case that the integral which describes the
vacuum energy\footnote{We could consider lagrangian instead of hamiltonian
with the same result.}
\be
\label{21}
\la H_{eff}\ra \sim 
\int_{0}^{\infty}\exp(-\frac{1}{t})
\frac{dt}{t(t+\frac{\la\Phi\ra}{\epsilon(\tau)}
)}\sim\epsilon\rightarrow 0 
\ee
goes to zero at small $\epsilon$.
As we mentioned above, the effect (\ref{21}) does not  crucially depend  on  
 our assumptions about the factorization properties for the condensates  
$\la \Phi^n \ra \sim \la \Phi \ra^n$ as neither on
our assumption of exact factorial dependence of the coefficients 
$c_n=n!$. Both of these effects presumably lead  (apart to $n!$) to some
mild $n-$ dependence which can be easily implemented
  into the formula
(\ref{21}) by introducing some smooth function $f(t)$ 
whose moments $\int f(t)t^{-n-2}\exp(-\frac{1}{t})dt$ exactly reproduce a
$n-$ dependence of the coefficients as well as of the condensates. 
If this function is mild enough, it will not destroy the relation
(\ref{21}), but might change some numerical coefficients.
Besides that, a condensate might have, along with singular part   
 proportional to
$\frac{\la\Phi\ra}{\epsilon(\tau)}$, a regular part as well
$\frac{\la\Phi\ra}{\epsilon(\tau)}+const.$
As can be seen from the representation (\ref{21})
this does not destroy the eq.(\ref{21}).

Few remarks are in order.
The vanishing of the vacuum energy is the consequence  of the
asymptotic nature of the effective lagrangian and the infrared properties 
of the VEVs. All others simplified assumptions which
have been made for technical reasons   do not affect the phenomenon.
Vanishing of the vacuum energy (\ref{21}) can be interpreted 
(after inflation, when all relevant  condensates
presumambly go to zero) as the vanishing of the cosmological constant,
{\bf the only relevant operator }in the Effective lagrangian
(all other terms  are marginal or unrelevant operators).

As our last remark, we would like to note that the 
strong infrared dependence of the vacuum condensate  
$ \la\Phi\ra $ is not a unique property of   de Sitter gravity.
Two-dimensional QCD with a large number of colors
also exhibits a strong
infrared dependence. In particular,
the so-called mixed vacuum condensates can be exactly 
calculated in this theory in the chiral
limit ($m_q\rightarrow 0$) and exhibit the following
dependence on the infrared parameter $m_q$ \cite{Chibisov}:
\be
\label{22}
 \frac{1}{2^n} \la  \bar{q}(ig\epsilon_{\lambda\sigma}
G_{\lambda\sigma}  \gamma_5)^nq \ra 
 =(-\frac{g^2\la\bar{q}q\ra}{2m_q})^n\la\bar{q}q\ra ,
 \ee
where $q$ is a quark field and $G_{\lambda\sigma}$ is a gluon
field of  $QCD_2(N=\infty)$. The chiral
condensate $\la\bar{q}q\ra$ in this theory can be calculated
exactly \cite{ARZ1}.  It does not vanish    
without contradicting   the Coleman theorem.
The very important feature of this formula: it diverges
in the chiral limit $m_q\rightarrow0$, where the parameter
$m_q$ plays the role of the infrared regulator of the theory.
Now , if we considered the 
asymptotic series  constructed from these condensates
\be
\label{23}
\sum_{n=0}^{n=\infty} (-)^nn! a_n
\la  \bar{q}(ig\epsilon_{\lambda\sigma}
G_{\lambda\sigma}  \gamma_5)^nq \ra
\sim\int_{0}^{\infty}\exp(-\frac{1}{t})
\frac{dtf(t)}{t(t+\frac{1}{m_q})}\sim m_q\rightarrow 0,
\ee
\be
a_n
\equiv\frac{1}{n!}\int_0^{\infty} f(t)t^{-n-2}\exp(-\frac{1}{t})dt\sim 1, \nonumber
\ee
we would get result of zero  for this series,
 in spite of the fact that each term 
on the left hand side   diverges
in the chiral limit and irrespective of the precise behavior of
the coefficients $a_n$ !
Of course this is   only a toy example 
which however can give us  a hint  of what might happen in   real Nature.

We close this section by noting that the vanishing of the vacuum energy
 in this scenario
{\bf  does not require any fine tuning of parameters.} Rather,
it is a very natural consequence of the asymptotic origin of the 
Effective lagrangian and of the infrared behavior of the VEVs.
The problem of naturality within an Effective lagrangian approach
has been discussed more than once. In the given context the 
cosmological constant problem has been discussed recently in 
\cite{Kaplan}with  the following 
 main conclusion: If a relevant operator
appears in the Effective field theory with a coefficient much less than a
typical scale without a symmetry reason, it should be taken 
as a warning for effective field theory dogma. 

We hope to have suggested  here a
 natural scenario for the vanishing of the coefficient for a
{\bf relevant operator} which is not based on    symmetry considerations.
We close this section with the following remark.
If this scenario   works (as we hope), it means first of all, that
all related   problems should be explained at the same time
within  the same approach. 
In particular, we expect \cite{BZ} that an inflationary scenario,
which is the most popular cosmology today,
can be understood   
  in     terms of the same physical
variables  within the same philosophy.

\section{Acknowledgments}
I am grateful to Robert Brandenberger,
 for the extremely valuable
lunch-discussions about gravity problems.
I also thank Sasha Polyakov for his useful
critical comments.

\end{document}